\documentstyle[psfig, twocolumn, floats, pra,aps]{revtex}

\begin{document}

\title{Slowing and cooling molecules and neutral atoms by \break time-varying
electric field gradients}

\author{Jason A. Maddi$^{1,2,*}$, Timothy P. Dinneen$^{1,\dag}$, and Harvey
Gould$^{1,\ddag}$}

\address{$^1$ Mail Stop 71-259, Lawrence Berkeley National Laboratory,
University of California,  Berkeley CA 94720.}

\address{$^2$ Department of Physics, University of California, Berkeley CA
94720.}

\date{September 15, 1999} \maketitle

\begin{abstract} A method of slowing, accelerating, cooling, and bunching
molecules and neutral atoms using time-varying electric field gradients is
demonstrated with  cesium atoms in a fountain. The effects are measured and
found to be in agreement with calculation. Time-varying electric field gradient
slowing and cooling is applicable to atoms that have large dipole
polarizabilities, including atoms that are not amenable to laser slowing and
cooling, to Rydberg atoms, and to molecules, especially polar molecules with
large electric dipole moments. The possible applications of this method include
slowing and cooling thermal beams of atoms and molecules, launching cold atoms
from a trap into a fountain, and measuring atomic dipole polarizabilities. \break
\break
\break
Pacs: 33.55.Be, 32.60.+i, 32.80.Pj, 33.80.Ps, 39.90.+d
\end{abstract}


\section{Introduction}

Time-invariant  electric field gradients have long been used to deflect beams
of molecules and neutral atoms. However, as we will show in this paper,
time-{\em varying} electric field gradients can be used to accelerate, slow,
cool, or bunch these same beams. We  demonstrate slowing, cooling, and bunching
of cold cesium atoms in a fountain, measure these effects and find good
agreement with calculation. The possible applications of the time-varying
electric field gradient technique include slowing and cooling thermal beams of
molecules and atoms, launching cold atoms from a trap into a fountain, beam
transport, and measuring atomic dipole polarizabilities.

\begin{figure}[tbh]
\centerline{\psfig{figure=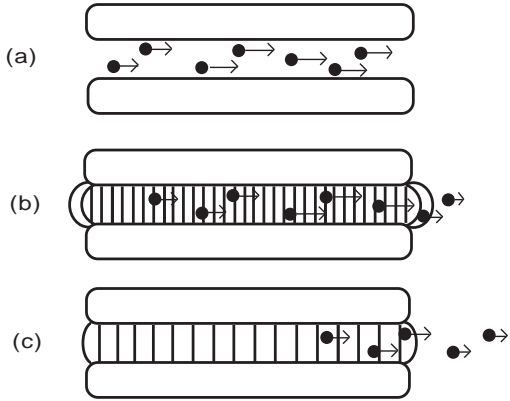,height=2.5in}} 
\caption{Schematic
diagram of longitudinal cooling with a time-varying electric field gradient.
The relative velocity of the particles is indicated by the length of their
arrows and the electric field strength by the density of field lines. A pulse
of molecules or neutral atoms expands as it enters the plates with the fastest
molecules or atoms in the lead(a). The same procedure as in Fig.~\ref{slowing
plates} is followed except that the electric field decreases over time. The
fastest, exiting first, lose more kinetic energy (b) than slower ones that exit
later when the field gradient is smaller (c). Alternatively, if the field is
turned on before all of the atoms have entered the electric field plates, the
slowest atoms are accelerated as they enter the plates. With sufficient field
strength, the relative velocities of the slow and fast molecules can be
interchanged and the pulse will rebunch. } \label{cooling plates} \end{figure}

The principle behind time-varying electric field gradient slowing is that an
electric field gradient exerts a force on an electric dipole (thus accelerating
or decelerating it) but a spatially uniform electric field, even if it is
time-varying, exerts no force on an electric dipole. Thus, an atom with an
induced electric dipole moment or a molecule with a `permanent' electric dipole
moment (with negative interaction energy in an electric field) will accelerate
when it enters an electric field and decelerate back to its original velocity
when it leaves the electric field. If we add a uniform electric field region
between the entrance and exit, as in a pair of parallel electric field plates
(Fig. \ref{slowing plates}), we can delay turning on the electric field until
the atom or molecule is in this uniform electric field. The atom or molecule
will not have accelerated entering the electric field plates but will
decelerate when it leaves the electric field, thus slowing. Longitudinal
cooling is achieved by applying a decreasing electric field, so that in a pulse
of atoms or molecules, the fastest ones, arriving first, experience the
greatest slowing (Fig. \ref{cooling plates}).
\begin{figure}[bh] 
\centerline{\psfig{figure=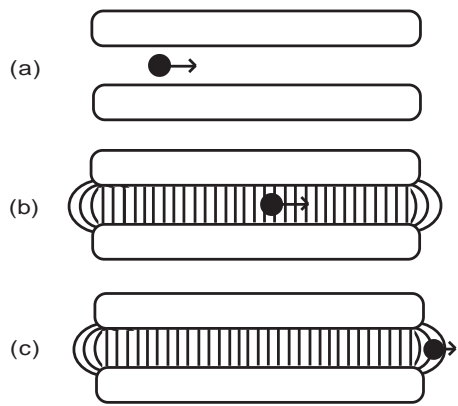,height=2.5in}}
\caption{Schematic diagram of slowing with a time-varying electric field
gradient. Molecules or neutral atoms enter the plates with the electric field
turned off (a). When they are between the plates, a voltage is applied
producing a spatially uniform electric field (b). They exit the plates,
traversing an electric field gradient where they are slowed (c). This process
may be repeated with additional sets of electric field plates.} \label{slowing
plates} \end{figure}

The time-varying electric field gradient technique can be useful for slowing
and cooling thermal beams of atoms with large dipole polarizabilities and polar
molecules with large electric dipole moments. Many atoms and most molecules are
not amenable to laser slowing and cooling, and presently few alternative
techniques\cite{buffer} exist. Slow molecules have application to molecular
beam spectroscopy, the study of chemical reactions, low energy collisions,
surface scattering, and trapping.

In molecular beam spectroscopy of long-lived states, slowing the molecules to
increase the transit time through the observation region can improve the
spectroscopic resolution, yielding better separation of spectroscopic features.
In thermal detection of molecular beams, the fraction of molecules in
long-lived vibrational states is measured by the extra energy that they
contribute to a thermal detector\cite{von Busch97}. The sensitivity of this
method can be increased by decreasing the kinetic energy of the beam.
Similarly, for control of chemical reaction pathways, reducing the kinetic energy a beam
of molecules to just above the reaction threshold energy may enhance the effect
of orientation or state preparation\cite{zare98}.

In the study of elastic scattering from surfaces, helium and a few other light
gasses are the most used projectiles because a small energy transfer to the
surface and narrow velocity distribution\cite{elastic} are essential.
Similarly, other forms of surface scattering\cite{scattering} also utilize
light atoms and light molecules (such as H$_2$, HD, CO, NO, and Ne). Heavier
atoms and molecules that have been slowed and cooled will also meet the
requirements of many surface scattering experiments. Thus, time-varying
electric field gradient slowing and cooling could make many more species
available for surface scattering experiments.

The trapping of molecules\cite{herschbach} can increase confinement,
observation, or interaction time by orders of magnitude, create high densities,
or allow the molecules to be cooled by evaporative cooling or other slow
cooling methods. With the possible exception of a toroidal storage ring
trap\cite{Katz}, slow cold molecules are a necessary prerequisite for all
proposed \cite{wing80,takekoshi,sekatskii,Friedrich,seideman} or
existing\cite{CaHVO} neutral molecule traps and would be an asset for a storage
ring\cite{Katz}. Time-varying electric field gradient slowing and cooling can
provide beams of slow, cold polar molecules in vacuum and is compatible with
all of these proposed or existing methods for trapping molecules.

The remainder of the paper is organized as follows. The interaction energies of
atoms and molecules in electric fields and the principles of slowing, cooling,
and bunching molecules and atoms with time-varying electric field gradients is
discussed in detail in Section~\ref{ame}. Our experiment is described and its
results are compared with calculation in Section~\ref{exp}. And finally, in
Section~\ref{applications}, we examine how the time-varying electric field
gradient method can be applied to slowing of thermal atoms and molecules,
measurements of atomic dipole polarizabilities, atom optics, and launching
atoms from traps.

\section{ Atoms and molecules in electric fields} \label{ame}

\subsection{Neutral atoms in electric fields}

Time-varying electric field gradient slowing utilizes the shift in an atom's
potential energy as it travels through a time- and spatially- varying electric
field. The effect of an electric field on an atom's potential energy is
described, to lowest order in the electric field, by the dipole polarizability
of the atom, defined as the ratio of the induced electric dipole moment to the
external electric field\cite{miller77,bonin94}. Although the dipole
polarizability is a tensor, the non-scalar terms are usually small, producing
only negligible variations in the polarizability of the different ground state
sublevels\cite{miller77,bonin94}, and do not affect the processes that we will
be discussing. Thus the induced dipole moment is, to a good approximation, a
scalar, and the potential energy, ${\cal E}$, is given, to lowest order in the
electric field, by ${\cal E} = -\alpha E^2/2$, where $E$ is the magnitude of
the electric field and $\alpha$ is the scalar dipole polarizability. Because
$\alpha$ is a scalar,  the potential energy depends only on the electric
field's magnitude and not its direction.

In a spatially varying electric field, the force is ${\bf F} =
(1/2)\alpha{\bf\nabla}(E^2)$, which, for all ground state atoms, is in the
direction of increasing electric field magnitude (strong field seeking). As
with any conservative potential, the change in an atom's kinetic energy, as
it travels between two points in space, is path independent and equal to the
change in potential energy between those points.

For example, a Cs atom traveling from a region, $E_i$ of no field, to a region
$E_f$ of $10^7$~V/m, gains kinetic energy, KE, by an amount
$\Delta\mbox{KE} = -\Delta{\cal E} = \alpha (E^2_f- E^2_i)/2 
= 3.3\times 10^{-25}\:\mbox{J} =
24\:\mbox{mK}$, where we have used the value of $6.63\times10^{-39}
\mbox{J/(V/m)}^2$ (or $59.6\times 10^{-24}$~cm$^3$) for the dipole
polarizability of Cs\cite{molof74,miller98}. Since we will be interested in
atomic beams from thermal sources (see Section IV), we will use energy units of
kelvin with the conversion $7.243\times 10^{22}$~K/J. The velocity of an atom
after traversing the potential is $v_f^2 = 2(\Delta\mbox{KE})/M+v_i^2$, where
$v_i$ and $v_f$ are the initial and final velocities, respectively, and $M$ is
the mass. For $v_i=0$, the final velocity for the example given above would be
1.70~m/s.

\subsection{Polar molecules in electric fields}

In addition to a dipole polarizability, polar molecules have an intrinsic
separation of charge that produces a dipole that can align with an external
electric field to yield a large net electric dipole moment \cite{books}. When
the interaction of the electric dipole moment, $d_e$, of a linear polar
molecule with an external electric field is large compared to the molecular
rotational energy, rotation is suppressed in favor of libration about the
direction of the electric field. The potential energy of the low-lying
rotational levels then approaches ${\cal E} = -d_e E$ and is always
negative\cite{von-meyenn70}. In a spatially varying electric field, the
resulting force is ${\bf F} = d_e {\bf\nabla}E$, which, as for ground state
atoms, is in the direction of increasing electric field magnitude (strong field
seeking).

As an example, consider cesium fluoride which has a very large dipole
moment\cite{lide} of $d_e=2.65\times10^{-29}$~J/(V/m) (or 7.88~Debye,
where 1~Debye $= 3.36\times10^{-30}$~J/(V/m)) and a small rotational
constant\cite{molecules} of $B_e =$ 0.27~K (or 0.188~cm$^{-1}$). In its lowest
angular momentum state ($J=0$), and traveling from a region of no field to
a region of $10^7$~V/m,  CsF  gains kinetic energy, KE,  by roughly the amount
$\Delta\mbox{KE} = -\Delta{\cal E} = d_e(E_f- E_i) = 2.65\times
10^{-22}\:\mbox{J} = 19\:\mbox{K}$. A more accurate value, calculated using the
formulas from Von Meyenn\cite{von-meyenn70}, is 16~K. This is about 640
times larger than for Cs, as discussed earlier.

As in the atomic case, the final velocity of the CsF molecule, after traversing
the potential, is $v_f^2 = 2(\Delta\mbox{KE})/M+v_i^2$, where $v_i$ and
$v_f$ are the initial and final velocities, respectively, and $M$ is the mass. For
$v_i=0$, the final velocity for the example given above, would be 45.7~m/s.
Equivalently, a 45.7~m/s CsF molecule traveling from a region of $10^7$~V/m
electric field, to a region of no field, would be slowed to rest.

\subsection{Slowing molecules and atoms}

A practical apparatus for slowing should have the electric field gradient
perpendicular to the field. Otherwise, a beam of molecules or atoms
traversing the electric field gradient is likely to strike one of the surfaces used to
form the electric field. A simple apparatus that meets this requirement is a
set of parallel electric field plates (Fig.~\ref{slowing plates}) attached to a
voltage source that can quickly be ramped from zero.

To operate a set of electric field plates as a time-varying electric field
gradient slowing apparatus, we do the following. A neutral atom in its
ground state enters the region between the electric field plates with no
field (the term atom also applies to clusters and molecules in strong field seeking states).
With the atom between the electric field plates, the voltage is turned on
producing a uniform electric field. The potential energy of the atom is lowered
by the electric field, but a spatially-uniform, time-varying electric
field does no work on a dipole, so there is no change in the kinetic energy of
the atom. As the atom exits the plates, passing through an electric field gradient, to a zero field region
outside, it gains potential energy and loses kinetic energy.
To accelerate an atom, the field is turned on before the atom enters the
electric field plates and is then turned off before the atom exits. The latter
arrangement can also be used to slow a weak field seeking molecule. 

The slowing process may be repeated by arranging a series of electric field
plates, each having a voltage applied once the atom has entered the uniform
electric field region. The energy change of the atom, traversing the
sequence,
is then cumulative. If a sufficient number of electric field plate sections are
assembled, it should be possible to slow a thermal beam of atoms to near rest.

This slowing process is analogous to, but the reverse of, the acceleration of
charged particles in linear accelerators\cite{accelerator} and cyclic 
accelerators \cite{lawrence30}, where charged particles
accelerate through a sequence of small voltage gradients. After each voltage
gradient, the charged particles drift through a time-varying, but spatially
uniform voltage, in which the voltage changes or reverses. This establishes a
new voltage gradient without requiring successively higher voltages.

The same slowing principle can also be applied using large magnetic field
gradients on atoms or paramagnetic molecules. However, it is more difficult to
switch strong magnetic fields.

\subsection{Cooling and bunching}

A decrease in the longitudinal velocity spread of the beam can be achieved by
applying an electric field, that decreases in time, to atoms that have been
arranged according to their velocity (Fig. \ref{cooling plates}). The first
atoms exiting the plates are slowed more than the atoms exiting at later
times,
when the electric field, and hence the electric field gradient, has decreased.
A beam of atoms will be ordered by velocity when a short pulse is allowed to
spread.

This is a form of cooling even though the initial and final velocity
distributions might not be Maxwell-Boltzmann. The process conserves phase space
(the area enclosed in a plot of the relative velocity of each particle versus
its relative position) and is analogous to debunching of a charged particle
beam in an accelerator. The process can also be reversed and used to bunch a
beam so that more atoms arrive at a selected point at the same time. Bunching
can reproduce or even compress the original longitudinal spatial distribution
of a pulse of atoms -- a useful technique for detecting weak signals.

\section{Experimental results} \label{exp}

\subsection{Experimental arrangement}

\begin{figure}[tbh] 
\centerline{\psfig{figure=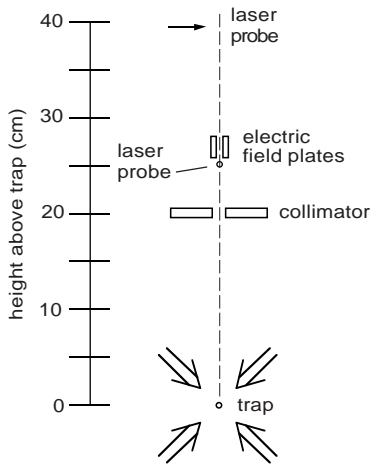,width=2.0in}}
\caption{Schematic of the apparatus used to test slowing and cooling. The
vertical dimension, the collimator, and plate spacing are to scale. The laser
probe beam below the electric field plates is parallel to the plate gap and the
probe above the plates is perpendicular to it.  Only four of the six trapping
lasers are shown.} \label{apparatus} \end{figure}

To test the principle of slowing and cooling with time-varying electric field
gradients, we slowed, cooled, and bunched packets of Cs atoms, initially
traveling 2~m/s, using a single set of electric field plates with fields of up
to $5\times 10^6$ V/m. The low initial velocity and large polarizability of Cs
made it easy to observe and measure the slowing (0.20~m/s at $5\times 10^6$
V/m), cooling, and bunching effects with a single electric field region. A
schematic of the apparatus is shown in Fig.~\ref{apparatus}.

Packets of Cs atoms were launched at rates of 0.25 Hz to 0.33 Hz from a
vapor-capture magneto-optic trap constructed along the lines described in
Ref.\cite{vapor,monroe92}. The laser beams that formed the x-y plane of the
trap were oriented at 45 degrees to the vertical. This made it easier to
perform measurements on the atoms after they had been launched. The trap
temperature, determined by observing the expansion of the Cs cloud, was about
30~$\mu$K. The trap laser system, which was stable and reliable, used an
external cavity grating configuration with piezo-electric tuning\cite{fox97},
and a Spectra Diode Labs model 5410-C diode laser with an antireflection
coating on the front facet. To launch the Cs atoms, a pair of acousto-optic
modulators blue (red) shifted the upward (downward) pointing laser beams by 5
MHz to form a moving molasses\cite{monroe92,fountain}.

The tower into which the atoms were launched extends 55~cm above the trapping
region. At 20~cm above the trap, a 1.3~cm aperture restricts the horizontal
dimensions of the packet. At 27~cm above the trap, a pair of stainless steel
electric field plates, each 2.2~cm tall by 1.7~cm wide, straddle the center
line. Each electric field plate is supported by a rod extending through a high
voltage vacuum feed-through mounted on a bellows. The bellows allowed us to
vary the spacing between the plates. Large gap spacings of 6~mm and 8~mm were
chosen to allow the maximum number of atoms through the plates, and to
minimize defocusing effects at the edges of the plates. See section~\ref{defocusing} for a
discussion of the defocusing effects.

Two high-voltage pulsed  power supplies, one positive and one negative,
were used to charge the electric field plates. The heart of each power supply is an
automobile ignition coil driven by a low current DC power supply that charges a
capacitor in series with the input of the ignition coil\cite{cclo}. Discharging
the capacitor supplies the input pulse to the coil. For cooling and bunching
experiments, a decaying voltage was produced by an RC circuit at the ignition
coil output. The components of this RC circuit are the high voltage coaxial
cable (about 100~pF/m) from the ignition coil output to the high voltage feed
through, and a resistor to ground. For slowing measurements, a high
resistance was chosen to make the time constant long compared to the transit time of the
atoms through the electric field plates.

For Cs atom time-of-flight velocity measurements, we formed probe laser beams
using a small fraction of the light from the trapping laser. One probe beam
passed 0.5~cm below the electric field plates and a second probe beam,
perpendicular to the electric field, passed 14~cm above the first. The probe
beam intensities were measured with photo diodes. The signal for the atoms
passing through a probe beam was the attenuation of the probe beam due to
scattering by the passing atoms.

The launched atoms arrived at the electric field plates in a packet 1.5~cm long
-- longer than the uniform region of the electric field plates. It was easier
to understand the results of slowing measurements if the packet of Cs atoms fit
entirely within the uniform region. The packet was trimmed, using the
lower probe beam, which deflected the atoms sufficiently, so that they were
not detected by the second probe. To accept only the center of the packet, the
laser was shifted out of resonance for a few milliseconds. We were thus able to
reduce the vertical size of the packet at the lower probe from 1.5~cm to about
0.3~cm. The arrival time of the atoms at the upper probe was measured relative
to the launch time. The stability of the launch was checked by periodically
measuring the arrival time at the lower probe.

\subsection{Slowing}

\begin{figure}[tbh] 
\centerline{\psfig{figure=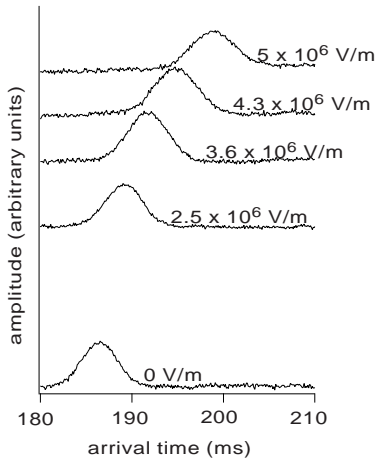,width=2.0in}}
\caption{Slowing of Cs atoms by time-varying electric field gradients. The time
of arrival at the upper probe is plotted for different electric fields used for
the slowing. The horizontal scale is the time elapsed from when the magnetic
field coils turn off for the launch. The vertical scale is the electric field
and superimposed on each field setting is the inverted absorption signal at the
upper probe.} \label{delay} \end{figure}

The time of arrival of the packet at the upper probe, as a function of the
applied electric field, is shown in Fig.~\ref{delay}. The electric field was
turned on after the Cs atoms entered the uniform field region of the plates,
and was kept nearly constant as the atoms exited the plates. Increasing the
electric field delayed the arrival of the Cs atoms at the upper probe. The
width of the packet increased because the packet had more time to spread.

\begin{figure}[tbh] 
\centerline{\psfig{figure=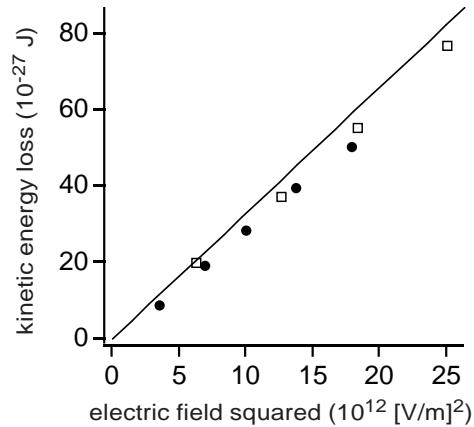,width=2.5in}}
\caption{Comparison of measured and calculated energy loss of Cs atoms
traversing a time-varying electric field gradient. The measurements were made
with 6 mm (shown as squares) and 8 mm (filled circles) plate spacings. The
statistical uncertainties are within the size of each point. The expected
energy loss, calculated from $\Delta\mbox{KE} = \alpha (E^2_f- E^2_i)/2$,
is shown as a line. The systematic
uncertainties are 2 percent for the Cs dipole polarizability and about 10
percent for our electric field. The observed differences between the measured
and expected energy loss is within these uncertainties.} \label{slow}
\end{figure}

For a quantitative measurement of the slowing, we calculated the loss in
kinetic energy, based on the increase in transit time, and plotted this
quantity as a function of the square of the electric field. This is shown in
Fig.~\ref{slow}. Two plate spacings, 6~mm and 8~mm, were used.  We compare
these data points with the expected energy loss, calculated from $\alpha E^2/2$
for $\alpha = 6.63\times10^{-39} \mbox{J/(V/m)}^2$ \cite{molof74}. The effect
is clearly quadratic in the electric field and is close to the predicted size.
The systematic error is consistent with the large uncertainty in our
measurement of the electric field.

\subsection{Cooling and compression}

To cool and bunch the Cs atoms we matched the decay time of the electric field
with the transit time of the atoms through the electric field gradient. In
addition, we used the full vertical size of the packet, which at the plates was
1.5~cm and, without cooling or bunching, was 2.5~cm at the upper probe. With
this short decay time,  we were able to utilize two methods to cool and bunch
the packets.   In the first,  the electric field was turned on once the whole
packet had entered the uniform region. Here, the faster atoms, exiting first,
were slowed more than the slower atoms, reducing the velocity spread.  In the
second method, we turned on the electric field once the faster atoms in the
leading edge of the packet had entered the uniform region of the field, but
while the slower atoms were still in the electric field gradient. The slower
atoms were accelerated into the plates, while the fastest atoms, already in
the uniform field region, were unaffected.  On exiting the plates,  the field
had decayed sufficiently so as not to affect the distribution.  However, with
plates of the proper length, additional cooling could be achieved, on exit,
using the first method. Bunching occurs as the packet evolves, if during the
process, the (initially) slow atoms are accelerated to a velocity greater than
that of the fast atoms.

\begin{figure}[tbh]
\centerline{\psfig{figure=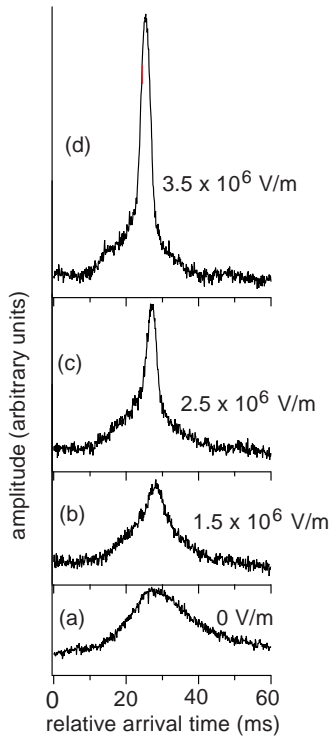,height=4.2in}} \caption{Cooling
and bunching of Cs atoms by time-varying electric field gradients. The signal
from the upper probe is shown as a function of the initial applied electric
field, which has a decay time of 5 ms. The narrowing of the base  in (b) - (d)
is due to cooling of the atoms. The narrow peak, especially in (c) and (d) is
due to bunching of the atoms.} \label{cooling}\end{figure}
As an example, Fig.~\ref{cooling} shows the arrival time
of the Cs atoms at  the upper probe as a function of electric field. The
electric field is turned on when roughly half of the packet reaches the uniform
field region. The rest of the packet is still in the electric field gradient.
The electric field decays with a RC time constant of 5~ms, and the packet
velocity at this point is roughly 2~m/s. The effects of cooling and then
compression in Fig.~\ref{cooling} are striking. The transit time of the atoms
through the upper probe is reduced from about 16~ms (FWHM) at zero field,
to 3~ms at $3.5\times 10^6$~V/m. With a packet velocity of 1.2~m/s at the upper
probe, the transit time corresponds to a packet length of about 3.6~mm,
reduced
from the original 1.5~cm at the electric field plates (2.5~cm at the upper probe).

\begin{figure*}[t] \centerline{\psfig{figure=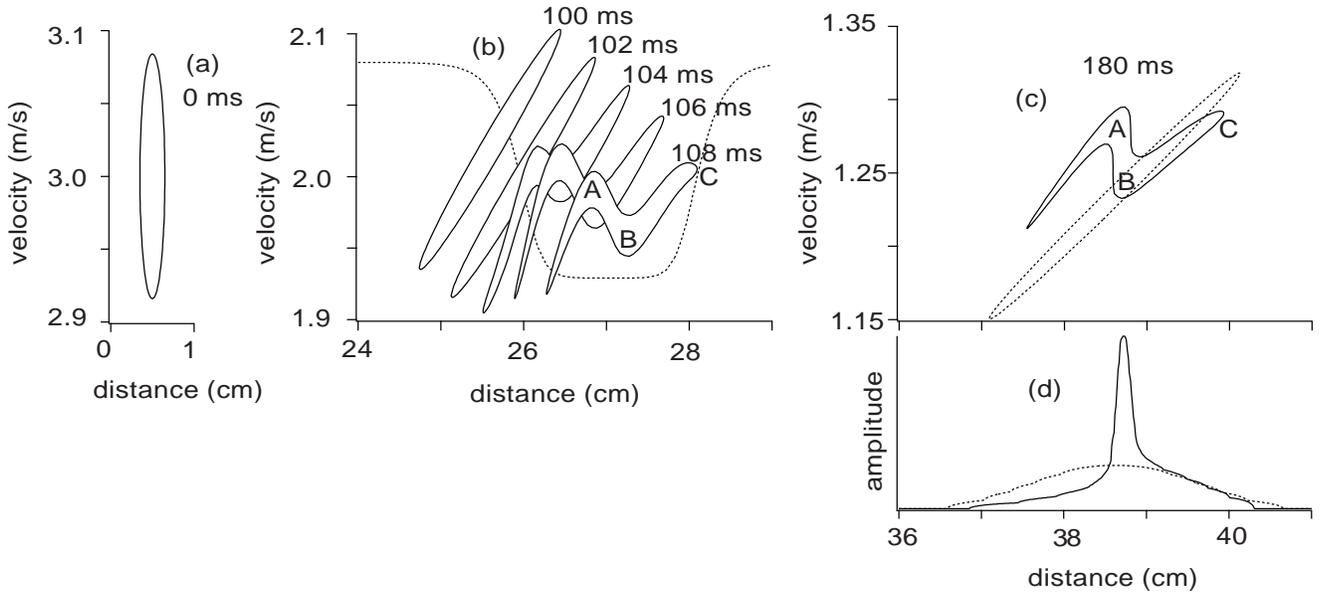,width=7in}}
\caption{Longitudinal phase space of the Cs atoms cooled and bunched in a
time-varying electric field gradient. The initial longitudinal phase space at
launch, shown in (a), is approximated by an ellipse with a 3~mm spatial spread
and a 0.15~m/s velocity spread.  At 100~ms after launch, (b), the packet has
spread to 1~cm and the velocity has decreased due to gravity. The electric
field, which has a decay constant of 5~ms, is turned on at 103~ms as the slower
atoms reach the electric field gradient at the entrance of the plates. The
dashed curve in (b) is the potential energy of the atoms due to the electric
field. From 103-106~ms the slow atoms are accelerated while the faster atoms
within the uniform field are unaffected. By 108~ms, the packet is past the
electric field gradient at the plate entrance and the electric field has
decayed, so there is little force on the atoms as they exit the plates. In (b),
the kink in the packet (marked A) is now at the same velocity as that of the
right-most portion (marked C).  The velocity spread of the packet has been
nearly halved.  As the packet time-evolves, (c), the spatial separation between
A and B does not change. In addition, the lowest velocity portion (marked B) is
retarded and, at 180~ms, lies at the same spatial coordinate as A. When this
occurs, we get a large peak in the absorption signal because the portion
between A and B arrives at the probe, as a bunch, only a few millimeters
wide (instead of the 2.5~cm for the uncompressed packet). The resulting calculated
density profile of the packet is shown in (d). The dashed curves in (c) and (d)
are the phase space and density profile, respectively, if no cooling had been
utilized. The densities in (d) incorporate an initial assumed Gaussian
phase-space distribution.} \label{ps1} \end{figure*}

To compare the experimental results with calculation, we modeled the time
evolution of the longitudinal phase space of the Cs atoms for the experimental
conditions in Fig.~\ref{cooling}. The results are shown in Fig.~\ref{ps1}. The
electric field along the center line between the plates, used to determine the
potential energy of the Cs atoms at each point, was calculated by a
two-dimensional finite-element analysis program. The resulting potential energy
is shown, superimposed on phase space diagrams, in Fig.~\ref{ps1}b.

\begin{figure}[tbh] 
\centerline{\psfig{figure=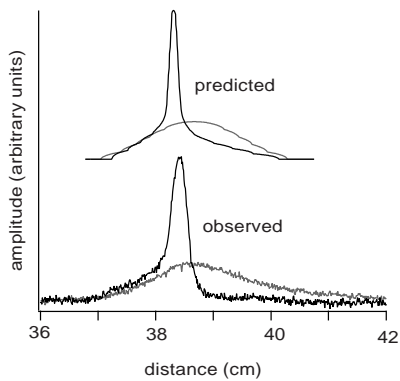,width=2.2in}}
\caption{Comparison of observed and predicted probe signals for cooling and
bunching of Cs atoms. The broad curves are for zero electric field and the
peaks are for an initial electric field of $3.5\times 10^6$~V/m that decays
with a 5~ms time constant. } \label{comparison} \end{figure}

In Fig.~\ref{comparison} we compare the observed Cs beam profile with the
calculation in Fig.~\ref{ps1}. The calculation, which is done in one dimension,
and assumes that the initial spatial distribution of atoms, is Gaussian. The
calculated spatial distribution
has been converted into time,  translated by about 5~ms, and scaled to align it to the
data. There is good agreement between experiment and calculation, except for a
small difference between the width of the calculated and observed peaks
(possibly due to the simple assumptions used in the calculation). We conclude
that one can make reliable calculations of the effects of time-varying electric
field gradients on the phase space evolution of atoms.

\subsection{Defocusing} \label{defocusing}

So far we have only discussed electric field configurations in one dimension.
For atoms on the midplane between two parallel plates there are no additional
forces. However, for atoms in the fringe field of the plates, and not on the
midplane, there is a force toward the nearest plate. The magnitude of this
force, which transversely defocuses the packet, depends on the shape of the
edges of the electric field plates. In general, any convex (concave) surface on
a field plate produces a local increase in the electric field gradient towards
(away from) the surface. It should be stressed that the change in kinetic
energy of the atoms is determined by conservation of energy. All atoms will
have their kinetic energy reduced, by the same amount, after exiting the
electric field, even though some may have slightly changed direction.

The defocusing effects can be minimized by using an electric field plate gap
that is large compared to the width of the beam of atoms. However, increasing
the gap increases the voltage needed to produce the same electric field, and
reduces the maximum electric field that can be sustained. For simplicity, in
this experiment, we chose a small gap-to-beam-width ratio and tolerated some
defocusing. However, there are more elaborate field configurations for which
our calculations show very small defocusing effects. In Fig.~\ref{splay}, we
compare one such set of electric field plates with field plates having a simple
parallel plate geometry.  For an atom slightly off the midplane, the transverse
force is reduced by about a factor of six compared to the simple plate
geometry. We will discuss details of focusing and defocusing in a future paper.

\section{Applications} \label{applications}

\subsection{Slowing thermal beams of atoms and molecules}

\subsubsection{Electric fields} \label{efield}

While electric fields of $10^7$~V/m or higher can be maintained by ordinary
metal electrodes with a small gap spacing,  much stronger fields can be
maintained by heated glass cathodes\cite{murray60}. Glass cathode systems have
been used to produce large Stark effects in beams of Cs\cite{marrus69} and Tl
atoms\cite{gould76}. A set of 75~cm long all-glass electric field plates have
operated at $4.5\times 10^7$~V/m\cite{gould76}. Short electric field plates
with a heated glass cathode at ground potential, and a metal anode have
sustained electric fields of $5\times 10^7$~V/m and higher\cite{marrus69}.

\subsubsection{Slowing ground state atoms}

Atoms that are of interest for slowing and cooling by time-varying electric
field gradients are those with large dipole polarizabilities that can not be
laser slowed and cooled. The dipole polarizabilities are largest in alkali
metals and alkali earths. However, actinides, lanthanides, and transition
elements near an alkali, also have polarizabilities\cite{miller98} above $1
\times 10^{-39} \mbox{J/(V/m)}^2$, compared to $6.63 \times 10^{-39}
\mbox{J/(V/m)}^2$ for Cs, which has the highest known ground state
polarizability.

As an example, we consider slowing a thermal beam of neutral americium (atomic
number 95) to near rest. With a polarizability\cite{miller98} of $2.59 \times
10^{-39} \mbox{J/(V/m)}^2$, each $5\times 10^7$~V/m slowing section will reduce
the kinetic energy by 0.23~K. It is impractical to slow atoms from the peak
of the velocity distribution at the 1500~K needed to form a beam of Am. However,
if one is willing to sacrifice intensity, the slower atoms, from the low
velocity tail of the thermal velocity distribution, are available. The low
velocity atoms from thermal distributions are often used to load magneto-optic
traps, either from a vapor inside the chamber\cite{vapor} (as we have
done), or
from an atomic beam\cite{beams,hulet}. Recently, Ghaffari et. al.\cite{hulet}
developed an atomic low-pass velocity filter that passes slow atoms and blocks
fast atoms.

For the Am velocity distribution tail at 1~K (8.4~m/s), only about five
time-varying electric field gradient slowing sections would be needed to bring
the atoms to near rest. The maximum energy spread that can be accepted by a
single section is about equal to the energy decrease in one section, which for
Am is about 0.23~K. Based upon a Maxwell-Boltzmann velocity distribution inside
a 1500~K effusive oven, with a thin orifice\cite{pauly}, roughly $10^{-7}$ of
the atoms in a beam will be in the energy range from 0.89~K to 1.1~K.  (
About $4\times 10^{-7}$ of the atoms in a beam will be within the energy range 3.60~K
to 3.83~K.)

The final velocity and the minimum length of the electric field plates will
determine repetition rate for slowing packets of atoms. At the last field plate
section, where the distance between pulses will be at a minimum, one pulse of
atoms must exit before the next pulse enters. The total number of
atoms that reach the end of the apparatus thus depends on the final velocity, the
length of the apparatus, the initial velocity, and on any focusing available. 

A
method for focusing strong field seeking atoms and molecules using alternating
electric field gradients\cite{focus},  has been applied to molecular beams\cite{focus2}.  
The range of beam
energies that can be accepted can also be increased by using a design that
cools over several electric field sections before slowing. We will discuss some
of these details in a future paper.

For clusters\cite{bonin94,bonin97}, the polarizability per atom of small
homonuclear alkali clusters is close to the atomic polarizability, and
decreases to a value of about 0.4 times the polarizability per atom for bulk
samples\cite{knight}. It should be possible to slow and cool clusters in the
same way as atoms.

We also note that the metastable states of noble gases have dipole
polarizabilities\cite{miller77} that are of order $10 \times 10^{-39}
\mbox{J/(V/m)}^2$. This would allow noble gas atoms in the metastable states to
be slowed and cooled, much the same way as ground state (Cs) atoms. They
do, however, have large tensor polarizabilities, that may permit effective slowing
and cooling of only a single angular momentum state.

\begin{figure*}[t] 
\centerline{\psfig{figure=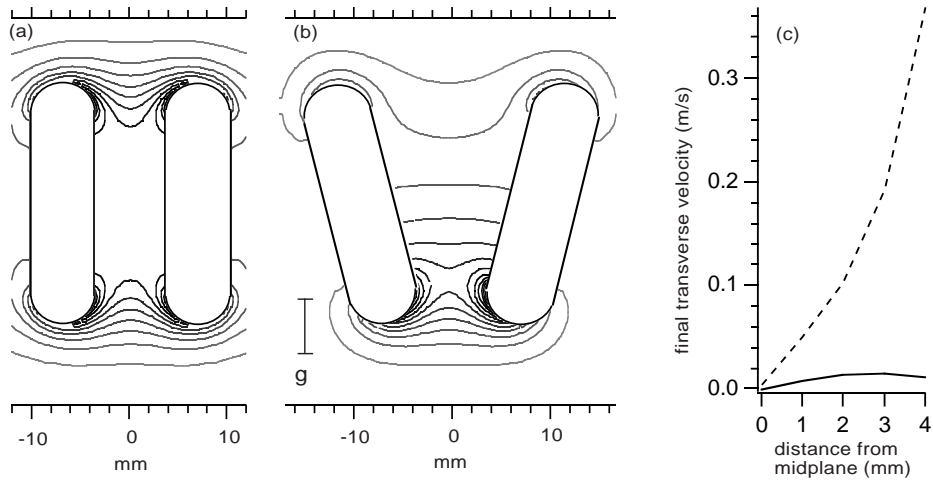,width=5in}}
\caption{Defocusing effects for simple parallel- and splayed-electric field
plates. In (a) and (b) the schematics of the two electric field plate
configurations are shown to the same scale, along with  the lines of constant
energy potential (${\cal E} = -(1/2)\alpha E^2$) for Cs from an applied voltage
of 40~kV. For comparison the interval corresponding to the force of gravity,
labeled g, is shown in (b). In the parallel plate configuration (a) used for
the measurements in this paper, the potential contours curve and atoms not on
the midplane acquire transverse velocities as they exit. In (b), the field
plates are splayed 15$^\circ$ from vertical producing a potential minimum just
above the bottom of the plates. On (b) the voltage would be applied once the
atom packet, traveling from below, has passed beyond the potential minimum.
The contours further up are much flatter than in (a), and the defocusing is
considerably less. The resulting transverse velocity for an initial
longitudinal velocity of 2~m/s is shown in (c).  The dashed line corresponds to
the simple parallel plates in (a) and the solid line corresponds to the splayed
plates in (b).} \label{splay} \end{figure*}

\subsubsection{Slowing Rydberg atoms}

The properties of Rydberg atoms\cite{gallagher}, atoms in states of very
high principle quantum number $n$, are similar for all elements and
include very large dipole polarizabilities\cite{miller77}, that can be either
positive or negative. This makes Rydberg states worth considering for
slowing atoms. Slowing of Rydberg atoms in inhomogeneous electric fields was proposed by Breeden and
Metcalf \cite{breeden-metcalf81}, who analyzed the case of time independent
inhomogeneous electric fields, and atoms in short-lived Rydberg levels.

To slow Rydberg atoms using time-varying electric field gradients, a number of
conditions must be met. The lifetimes of the states must be long enough to pass
through the apparatus and the electric field must neither quench the state, nor
or ionize the atom, but should still be large enough to produce significant
slowing. The lifetime of a state in a quasi-hydrogenic Rydberg atom, with
principle quantum number $n$, and angular momentum $l$, has been
calculated by Chang \cite{chang85}, who finds for high angular momentum, $\tau = 93 n^3 (l
+0.5)^2$, where $\tau$ is the lifetime in ps. For $n=30$, the lifetimes are
about 2.2~ms for $l=29$ and about 1.1~ms for $l=20$. An external electric
field mixes different values of $l$ having the same z component of angular
momentum $m$. Thus, only the sublevels with large values of m will have
unquenched lifetimes, since the lower m states mix with low $l$ states that
have much shorter lifetimes.

The critical electric field for ionizing a Rydberg atom is given classically as
$E_{cr} = 1/(16n^4)$, where $E_{cr}$ is in atomic units (of $5.14\times
10^{11}$~V/m). High m states are more circular and thus require a higher field
to ionize. The Stark effect also modifies the critical field \cite{gallagher}
and for blue shifted levels the critical field may be closer to $1/(12n^4)$.

The change in energy levels with the electric field has been calculated by
Bethe and Salpeter \cite{bethe-salpeter}. For a hydrogenic case they find, in
atomic units (1~au $= 1.0973\times 10^7$~m$^{-1}$), $W = -0.5 n^2 + 1.5 E n
(n_1-n_2) -E^2 n^4 [17 n^2 -3 (n_1-n_2)^2 -9 m^2 +19]/16$, where $n_1$
and $n_2$ are parabolic quantum numbers that satisfy the equation $n= n_1 + n_2 +|m| +1$.
Setting $E$ equal to $1/(12n^4)$, we find that for $n=30$, $l=20$ 
the energy change
at the maximum field, that does not ionize the atom, is about 5.6~K  or 7.6~K,
depending on whether the red or blue shifted level is chosen. For the circular
state, $|m| = n-1$ and $n_1=n_2 = 0$, the electric field can not mix states
from the same $n$ or lower $n$  and the shift is much smaller. For $n=30$,
$l=29$, the maximum energy change before ionization is about 0.66~K.

\subsubsection{Slowing polar molecules}

Time-varying electric field gradient slowing and cooling of polar molecules
with large electric dipole moments can be very efficient. As examples we
consider two linear diatomic molecules with large dipole moments: cesium
fluoride, which has a small rotational constant, and lithium hydride, which has
a very large rotational constant.

\begin{figure}[tbh]
\centerline{\psfig{figure=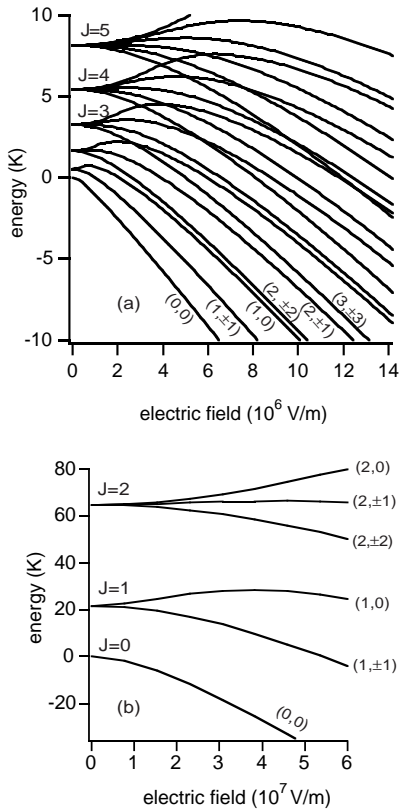,height=4.5in}} \caption{Energy
levels of the low rotation levels of CsF (a) and LiH (b) in electric fields.
The calculation is based on the equations in von
Meyenn\protect\cite{von-meyenn70} who uses a rigid rotor model. The energy is
in units of kelvins (1~K $=$ 0.69~cm$^{-1}$) Note the different scales for the
electric field in (a) and (b) .} \label{molenergy} \end{figure}

As discussed in section~\ref{ame}, a rigid rotor model
calculation\cite{von-meyenn70} of CsF, in its lowest rotational state
($J=m=0$), shows that CsF would lose about 16~K of energy exiting each
$10^7$~V/m electric field section. The next few higher rotational levels (Fig. \ref{molenergy}a)
would also experience large changes in kinetic energy and could be efficiently
slowed and cooled. With a $5 \times 10^7$~V/m electric field, the change in
kinetic energy (see Fig. \ref{molenergy}a), for the lowest rotational level,
would be about 89~K, or equivalently a molecule traveling 98~m/s could be
brought to rest. Longitudinal cooling, of about 89~K, could also be achieved
in a single time-varying electric field gradient section.

In a thermal beam of CsF formed\cite{hughes47} at 850~K, about one-half
percent of the molecules have a kinetic energy of 89~K or less. A disadvantage of using
a thermal beam of molecules with a small rotational constant is that only a
very small fraction of the molecules are found in any single rotational state.
Only about one in 15,000 CsF molecules are found in the $J=m=0$ state in a
thermal molecular beam at this temperature\cite{hughes47}.

An alternative approach is to form a supersonic beam\cite{morse96} of CsF. The
supersonic beam has a much lower internal temperature, which greatly
increases the population of low rotational levels, and it is a very directional
beam with a narrow velocity distribution. However, the beam velocity of a supersonic
source is higher than the most probable velocity from an effusive source of the
same temperature and has no low velocity tail. Slowing a supersonic CsF beam
would require approximately twenty electric field sections -- still very
doable.

Lithium hydride and other metal hydrides have both very large rotational
constants and large dipole moments. For LiH, $B_e =$ 11~K (7.5~cm$^{-1}$) and
$d_e=2.0\times10^{-29}$~J/(V/m) (5.9~Debye)\cite{molecules,lide}. Even at  $5
\times 10^7$~V/m, the electric field does not completely suppress rotation
(Fig. \ref{molenergy}b), and for the $m=0$, $J\ne 0$ states, the interaction
energy is both large and positive. As shown in Fig. \ref{molenergy}b, a LiH
molecule in the $J=1$, $m=0$ state {\em entering} an electric field of 
$3.5 \times
10^7$~V/m would lose about 7~K of energy. And since the rotation constant is
large, a significant fraction of the molecules in a thermal beam are in low
rotation states. Approximately $2 \times 10^{-5}$ of the LiH molecules in a
beam from a1200~K effusive oven will have an energy of 7~K or less. 
Those in the $J=1$, $m=0$ state could be slowed to rest with a single electric
field section.

Lithium hydride in the low-lying $m=0$, $J\ne 0$ states and other molecular
states that have positive interaction energies are weak field seeking and can
be transversely focused by static multipole electric fields\cite{bernstein}.
(For focusing strong field seeking states see Ref.\cite{focus,focus2}).
It may also be possible to trap weak field seeking states in electric
field traps \cite{Katz,wing80,takekoshi,sekatskii,seideman} or a laser
trap\cite{Friedrich}. An electric field trap could, in principle, be up to
7.5~K deep for the $J=1$, $m=0$ state of LiH, depending on the 
electric field that
can be sustained in the trap geometry.

The $J=1$, $m=0$ and $J=2$, $m=0$ rotational levels of CsF have 
negative interaction
energies at strong electric fields, but positive interaction energies at weaker
fields (Fig.\ref{molenergy}a). Thus, CsF in these states could be efficiently
slowed using strong electric field gradients, then focused and trapped in a
weaker electric field. The change in kinetic energy exiting a $5 \times
10^7$~V/m electric field would be 82~K (74~K) for the $J=1$, $m=0$ 
($J=2$, $m=0$)
state, while in an electric field of $2.5 \times 10^6$~V/m  ($4 \times
10^6$~V/m) the $J=1$, $m=0$ ($J=2$, $m=0$) state would be weak field 
seeking with an
interaction energy of 0.5~K (1~K). Alternatively, CsF could be slowed in the
$J=1$, $m=1$ state and then, in a weak electric field, 
transferred to the $J=1$, $m=0$ state, by an rf transition\cite{hughes47}.

\subsection{Application to cold atoms}

\subsubsection{Measurement of the ground state dipole polarizability of atoms}

Fig.~\ref{slow} demonstrates the sensitivity of time-varying electric field
gradient slowing to the static dipole polarizability. Each data point in this
figure represents less than 300 seconds of counting. Key quantities in a
time-varying electric field gradient measurement of dipole polarizabilities are
the electric field strength, the beam velocities, and the electric field
profile at the plate exit and/or entrance. The field profile is needed to
determine the drift lengths before and after slowing for a time-of- flight velocity measurement. We
expect that with a moderate effort, polarizability measurements on alkali,
alkaline earth and other slow, cold atoms can be made to an accuracy of a few
parts per thousand\cite{cesium}.

Although dipole polarizability is related to many important physical and
chemical properties \cite{miller98}, the ground state dipole polarizability has
been measured in fewer than 20 percent of the known elements\cite{miller98}.
And of these, only for the noble gasses and sodium \cite{ekstrom95} has an
accuracy of one percent been surpassed\cite{miller98}. The traditional method
for measuring the static dipole polarizabilities of condensible atoms is the
elegant electric-magnetic field gradient balance technique (E-H balance)
\cite{miller77,salop61}, which uses thermal beams of atoms. Slow, cold
atoms would also allow a significantly improved accuracy for E-H balance
and other deflection based methods. However, we anticipate that time-varying electric
field gradient slowing (or acceleration) will be easier to perform and poses
fewer challenges to understanding the distributions of electric field and
atoms.

\subsubsection{Beam transport}

Inhomogeneous magnetic fields are used for transverse focusing of laser-cooled
atom packets\cite{kaenders96}, and Cornell, Monroe, and Wieman\cite{cornell91}
have used time varying inhomogeneous magnetic fields to radially and axially
focus atoms being transferred between traps. Time-varying electric field
gradients are a useful compliment to these atom optic elements because they
are insensitive to the magnetic or hyperfine substates, and when edge effects 
are small or absent, they control only in the longitudinal direction.

One possible application of the time-varying electric field gradient to beam
transport, is to longitudinally spread atoms in a cold Cs atom atomic clock, to
reduce collisional frequency shifts\cite{gibble}. In such a clock, the atoms
would be spread before they passed through the first rf region (or the first
passage through the single rf region in a fountain clock) and, if necessary, 
to sharpen the detection signal,
rebunched after they passed through the second rf region (after their return
through the rf region in a fountain clock) . The
magnetic fields associated with turning on and off the electric field can be
made small so as not to influence the magnetic shielding environment of the
clock.

\subsubsection{Launching atoms}

If a set of electric field plates is turned on near a cloud of cold confined
atoms in the ground state, the atoms will accelerate into the plates. Turning
off the field when the atoms are in the uniform field region then allows the
atoms to exit the plates with a net velocity. Cesium atoms entering an electric
field of $4 \times 10^7$~V/m will accelerate from rest to 6.8~m/s.

The electric field plates can be positioned to launch atoms in any direction.
To launch horizontally, a vertical gap will provide a fringe field at the
bottom that can help keep the atoms from falling out of the plates. To launch
vertically, the electric field gradient at the initial location of the atoms
needs to be large enough to overcome gravity. In microgravity, the initial
acceleration needs only to be large enough that the cloud of atoms does not
expand beyond the dimensions of the electric field plate gap (to prevent the
loss of atoms). It would also be easy to
vary the launch velocity and direction. 
One possible arrangement of electric field plates is to invert
the plate configuration shown in Fig.~\ref{splay}.

\section{Acknowledgements}

We are grateful to Douglas McColm for many stimulating and fruitful discussions
and for his help in clarifying several key concepts. We thank Timothy Page,
Andrew Ulmer, Christopher Norris,  Karen Street, and Otto Bischof for
assistance in constructing the apparatus, and thank C.C. Lo for developing a
very functional and cost-effective time-varying high voltage power supply. One
of us (JM) thanks the Environment, Health, and Safety Division at LBNL, and
especially Rick Donahue and Roberto Morelli, for help with computing resources,
and one of us (HG) thanks Alan Ramsey for timely inspiration. This work was
supported by the Director, Office of Science, Office of Basic Energy Sciences,
of the U.S. Department of Energy under Contract No. DE-AC03-76SF00098. One of
us (JM) is partially supported by a National Science Foundation Graduate
Fellowship.


\begin{thebibliography}{99}
\bibitem{cesium} An electric field gradient slowing measurement of the Cs
dipole polarizability is currently underway in our laboratory.
\end{thebibliography}
\end{document}